\documentclass[10pt,twocolumn,english,floatfix]{revtex4-1}

\usepackage{amsthm}
\usepackage{dsfont}
\usepackage[T1]{fontenc}
\usepackage[latin9]{inputenc}
\usepackage{geometry}
\usepackage{amsmath}
\usepackage{amssymb}
\usepackage{graphicx}
\usepackage{color}

\newcommand{\bR}{\mathbb R}
\newcommand{\bC}{\mathbb C}

\makeatletter
\usepackage{babel}

\makeatother

\begin{document}

\title{%\begin{multicols}{2} % open this to use the two-column format
Certification and the Potential Energy Landscape}
% Certifying the Potential Energy Landscape - II}

\author{Dhagash Mehta}
\email{dbmehta@ncsu.edu}
\affiliation{Dept of Mathematics, North Carolina State University, Raleigh, NC 27695, USA;\\
Dept of Chemistry, The University of Cambridge, Cambridge, CB2 1EW, UK.}
\author{Jonathan D.~Hauenstein}
\email{hauenstein@ncsu.edu}
\affiliation{Dept of Mathematics, North Carolina State University, Raleigh, NC 27695, USA.}
\author{David J.~Wales}
\email{dw34@cam.ac.uk}
\affiliation{Dept of Chemistry, The University of Cambridge, Cambridge, CB2 1EW, UK.}

%\author{Dhagash Mehta$^{a}$\thanks{This is footnote}, Jonathan D.~Hauenstein$^{b}$, David J.~Wales${^{c}}$\\
% \mbox{}\\
% %\mbox{}\\
%$^{a}${\em Dept of Physics, Syracuse University, Syracuse, NY 13244, USA.}\\
%$^{b}${\em Dept of Mathematics, North Carolina State University, Raleigh, NC 27695, USA.}\\
%$^{c}${\em Dept of Chemistry, The University of Cambridge, Cambridge, CB2 1EW, UK.}}

\begin{abstract} \noindent Typically, there is no guarantee that a numerical approximation obtained using
standard nonlinear equation solvers is indeed an actual solution, meaning that it lies in the quadratic convergence basin.
Instead, it may lie only in the linear convergence basin, or even in a chaotic region, and hence not converge to
the corresponding stationary point when further optimization is attempted. In some cases, these non-solutions could be misleading.
Proving that a numerical approximation will quadratically converge to a stationary point is termed \textit{certification}.
In this report, we provide details of how Smale's $\alpha$-theory can be used to certify numerically obtained stationary points of a potential energy landscape, providing a \textit{mathematical proof} that the numerical approximation does indeed
correspond to an actual stationary point, independent of the precision employed.
\end{abstract}

\maketitle

\section{Introduction}

Given a potential $V({\bf x})$, with ${\bf x}=(x_{1},\dots,x_{n})$, the surface defined by $V({\bf x})$
is called the potential energy landscape (PEL) of the given system \cite{Wales:04}.
The special points of a PEL, defined by the solutions of the equations
$\partial V({\bf x})/\partial x_{i}=0$ for $i=1,\dots,n$, provide important
information about the PEL. These special points, called critical points or
stationary points (SPs) of the PEL, can be further classified according to the number of negative eigenvalues of the Hessian matrix,
$H_{i,j}=\partial^{2}V(x)/\partial x_{i}\partial x_{j}$. The SPs at which ${\bf H}$ is positive (negative) definite are called
minima (maxima) of the PEL and the SPs at which ${\bf H}$ has exactly $I$ negative eigenvalue are called saddles
of index $I$.  SPs at which ${\bf H}$ has at least one zero eigenvalue, after removing the global
symmetries from the system corresponding to overall translation and rotation,
are called singular SPs or non-Morse points.

The SPs of the PEL can be employed to calculate or estimate certain physical quantities of interest. 
A variety of techniques have been deveoped within the
framework of potential energy landscape theory \cite{Wales:04,RevModPhys.80.167},
with applications to many-body systems as diverse as
metallic clusters, biomolecules and their folding transitions, and glass formers.

Except for rare examples, like the one-dimensional XY model \cite{Mehta:2010pe}, it is not usually possible to obtain the
SPs analytically because solving the nonlinear stationary equations can be an extremely difficult task. Hence, one has to rely upon numerical methods.
When a numerical method finds a solution of a given system, it essentially means that it has found a numerical approximation of an exact solution.
After achieving a numerical approximate, one can heuristically validate it by either monitoring iterations of Newton's method or by substituting the approximations
into the equations to see if they are satisfied up to a chosen tolerance. Usually, such a validation works well in practice.
However, as will be clear from the examples provided below, such heuristic approaches do not guarantee that the numerical approximation will indeed converge quadratically to the
associated solutions using arbitrary precision. More specifically, even if a numerical approximation is heuristically validated, it could turn out to be a nonsolution at higher precision, or Newton iterations may have unpredictable behavior, such as attracting cycles and chaos, when applied to points that are
not in a basin of attraction \cite{mezey81b,mezey87,wales92,wales93d,Asenjo13} of~some~solution.

We note that if the given system is a set of polynomial equations, then one can use numerical polynomial homotopy continuation
\cite{Mehta:2009,Mehta:2009zv,Mehta:2011xs,Mehta:2011wj,Kastner:2011zz,Maniatis:2012ex,Mehta:2012wk,Hughes:2012hg,Mehta:2012qr,MartinezPedrera:2012rs,
He:2013yk,Mehta:2013fza,Greene:2013ida,SW:05}
to compute all the isolated solutions (see e.g.~\cite{KowalskiJ98,ThomH08,pielaks89} for some related approaches). Briefly, the method works as follows: first, one determines
an upper bound on the number of isolated complex solutions of the given system. The highest upper bound on the number of solutions is the so-called Classical Bezout bound, which is the product of the degree of each polynomial equation of the system, but there are several other tighter upper bounds available for structured systems. Then, one constructs another system that has exactly the same number of solutions as the upper bound, such that the new system is easy to solve. Finally, one uses continuation to track each solution
of the new system to the original one.  Some of the paths may diverge to infinity, which will happen whenever the upper bound is larger
than the true number of solutions.  The paths that converge will tend to solutions of the original system
in appropriate limits.
This method is quite different from conventional numerical approaches, in that it is guaranteed to find
all solutions, in principle. However, due to the numerical computations used with this method in path tracking, in the end, only numerical approximations are obtained, and hence the above mentioned difficulties may also arise.  The goal of this paper is to develop a rigorous way to validate the numerical approximates of the SPs, independent of the numerical method used to obtain them.

In the mathematics literature, proving that a given numerical approximation will converge quadratically
to the nearby associated solution using arbitrary precision is called \textit{certification}. It is well known that quadratic convergence doubles the number
of correct digits after each iteration. Hence the associated solution can be approximated to a given accuracy quite efficiently after a certain number of Newton iterations.
Smale and others, in the 1980's, developed a method that certifies a numerical approximation as an actual solution of the
system \cite{BCSS}. The method is now known as Smale's $\alpha$-theory. Interestingly, it turned out that the certification could be done via computing three numbers from a given numerical approximate: for a given system of equations $f = 0$
and a given point $x^*$, one computes two numbers $\beta(f,x^*)$ and $\gamma(f,x^*$), which guarantee that Newton's method starting from $x^*$ will quadratically converge to a solution
of $f = 0$ if the number $\alpha(f,x^*) = \beta(f,x^*) \gamma(f,x^*)$ is
less than $\left(13-3\sqrt{17}\right)/4\thickapprox0.157671$.
Applying this certification scheme
ensures that our numerical solutions
are good enough so that more accurate approximations
can be obtained easily and efficiently \cite{Mehta:2013zia}.

In this paper, we first give details of Smale's $\alpha$-theory in the context of the PELs in Section \ref{Sec_Smales_theory}. Then, by providing examples in Section \ref{Sec_Examples}, we show how numerical approximates may turn out to be non-solutions even in seemingly simple situations. In passing, we certify the solutions of the well-known examples of the Wilkinson polynomial under a small perturbation and the roots of the Chebyshyv polynomials of the first kind for the first time. In Section \ref{sec:XY_model}, we consider a more physically relevant potential,
i.e., the two-dimensional XY model without disorder.
We certify all the known SPs of this model and provide a guide for conventional numerical methods.
Section \ref{Sec_Conclusions} provides an outlook and conclusions.

\section{Smale's $\alpha$-Theory}\label{Sec_Smales_theory}

In this section, we describe Smale's $\alpha$-theory following Ref.~\cite{2010arXiv1011.1091H}.
We restrict ourselves to square systems, i.e., systems that have the same
number of equations as variables, since SPs of a PEL satisfy a square system of equations.  Smale's $\alpha$-theory
is usually used to certify complex solutions for systems of analytic functions, so we start by describing this
approach \cite{BCSS}.
We then discuss the certification of real solutions separately. Finally, we will discuss
$\alpha$-theory applied to polynomial systems and systems involving exponentials and trigonometric functions.

We start by considering a system $f$ of $n$ multivariate analytic equations in $n$ variables.
We denote the set of solutions of $f = 0$ as $\mathcal{V}(f):=\{{\bf z}\in\mathbb{C}^{n}|f({\bf z})=0\}$
and the Jacobian of $f$ at ${\bf x}$ as $J_{f}({\bf x})$.
Consider the Newton iteration of $f$ starting at ${\bf x}$ defined by
\begin{equation}
N_{f}({\bf x}):=\begin{cases}
{\bf x}-J_{f}({\bf x})^{-1}f({\bf x}), & \mbox{if }J_{f}({\bf x})\mbox{ is invertible,}\\
{\bf x ,} & \mbox{otherwise.}
\end{cases}
\end{equation}
For $k\geq1$, the $k$-th Newton iteration is simply
\begin{equation}
N_{f}^{k}({\bf x}):=\underbrace{N_{f}\circ\dots\circ N_{f}}_{\hbox{$k$ times}}({\bf x}).
\end{equation}

Now, a point ${\bf x}\in\mathbb{C}^{n}$ is called an {\em approximate solution}
of $f$ with {\em associated solution} ${\bf z}\in\mathcal{V}(f)$ if, for each $k\geq1$,
\begin{equation}
\left\|N_{f}^{k}({\bf x})-{\bf z}\right\|\leq\left(\frac{1}{2}\right)^{2^{k}-1}\left\|{\bf x}-{\bf z}\right\|,
\end{equation}
where $\|\cdot\|$ is the standard Euclidean norm on $\mathbb{C}^{n}$, i.e., ${\bf x}$ is an approximate solution to $f$ if it is in the
quadratic convergence basin defined by Newton's method of some solution~${\bf z}$.
The following theorem provides a sufficient condition for proving that a given point
is an approximate solution without knowledge~about~${\bf z}$.

\textbf{Theorem:} If $\alpha(f,{\bf x})<\left(13-3\sqrt{17}\right)/4$
for a square analytic system $f$ and point ${\bf x}$ such that $J_{f}({\bf x})^{-1}$ exists,
then ${\bf x}$ is an approximate solution to $f$, where
\begin{equation}
\begin{array}{cl}
\alpha(f,{\bf x}):= & \beta(f,{\bf x}) \gamma(f,{\bf x}),\\
\noalign{\smallskip}
\beta(f,{\bf x}):= & \|J_{f}({\bf x})^{-1} f({\bf x})\|, \quad\ \mbox{and}\\
\noalign{\smallskip}
\gamma(f,{\bf x}):= & \underset{k\geq2}{\mbox{sup}}
\left\|\frac{\displaystyle J_{f}({\bf x})^{-1}D^k {f} ({\bf x})}{\displaystyle k!}\right\|^{1/(k-1)}.
\end{array}
\end{equation}
If ${\bf x}$ is an approximate solution of $f$, then $\|{\bf x}-{\bf z}\|\leq2\beta(f,{\bf x})$,
where ${\bf z}\in\mathcal{V}(f)$ is the associated solution to ${\bf x}$.
Moreover, in $\gamma(f,{\bf x})$, the term $D^k f({\bf x})$ is the symmetric tensor
whose components are the partial derivatives of $f$ of order $k$.
Finally, for convenience, one can remove the condition on $J_{f}({\bf x})$ by
defining $\alpha$, $\beta$, and $\gamma$ appropriately.
If ${\bf x}\in\mathcal{V}(f)$ such that $J_{f}({\bf x})$ is not invertible,
define $\alpha(f,{\bf x}):=0$, $\beta(f,{\bf x}):=0$ and $\gamma(f,{\bf x}):=\infty$.
If ${\bf x}\notin\mathcal{V}(f)$ such that $J_{f}({\bf x})$ is not invertible,
then $\alpha(f,{\bf x})$, $\beta(f,{\bf x})$ and $\gamma(f,{\bf x})$ are taken as $\infty$.

Since this theorem provides a sufficient condition
for a point to be an approximate solution, the set of
{\em certifiable} approximate solutions is generally
much smaller than the set of approximate solutions, as demonstrated in
Figures~\ref{Fig:Z4-1} and~\ref{Fig:Cheby}.
However, if it is a true approximate solution, then a few Newton iterations usually generate a point that can be certified.

\noindent {\bf Distinct Complex Solutions}

Given two approximate solutions ${\bf x}_1$ and ${\bf x}_2$, one often
needs to verify that the corresponding associated solutions ${\bf z}_1$ and ${\bf z}_2$
are distinct.  One way to check this condition uses the triangle inequality together with
$\|{\bf x}_i-{\bf z}_i\|\leq2\beta(f,{\bf x}_i)$.

\subsection{Special Nonlinear Systems}\label{Sec:SpecialNonlinear}

For arbitrary analytic systems, $\gamma(f,{\bf x})$ as described in the above
theorem may be difficult to compute or bound above.  When $f$ is polynomial,
$\gamma$ is defined as a maximum over finitely many terms and thus can be computed
in theory.  Often, however, such as in the software package {\tt alphaCertified} \cite{2010arXiv1011.1091H},
$\gamma(f,{\bf x})$ is bounded above via Proposition~8 of \cite[\S~I-3]{Bezout1},
which depends on the degrees and coefficients of the polynomials in $f$, $\|x\|$,
and $J_f({\bf x})$, which we now briefly summarize.

Let ${\bf x}\in\bC^n$ and $g$ be a polynomial in $n$ variables of degree~$d$.
Define
$$\|{\bf x}\|_1^2 = 1 + \|{\bf x}\|^2 = 1 + \sum_{i=1}^n |x_i|^2$$
and, by writing $g({\bf x}) = \sum_{|\rho|\leq d} a_\rho {\bf x}^\rho$, define
$$\|g\|^2 = \sum_{|\rho|\leq d} \rho!\, (d - |\rho|)!\, |a_\rho|^2.$$
For a system $f$ of $n$ polynomials in $n$ variables, let
$$\|f\|^2 = \sum_{i=1}^n \|f_i\|^2.$$
If $J_f({\bf x})$ is invertible, define
$$\mu(f,{\bf x}) = \|f\|\times\|J_f({\bf x})^{-1}\Delta_{(d)}(\bf x)\|$$
where $d = (d_1,\dots,d_n)$ with $d_i = \deg f_i$ and
$$\Delta_{(d)}(\bf x) = \left[\begin{array}{ccc} d_1^{1/2} \|{\bf x}\|_1^{d_1-1} & & \\
& \ddots & \\ & & d_n^{1/2}\|{\bf x}\|^{d_n-1}\end{array}\right].$$

\textbf{Proposition:} If $f$ is a polynomial system
with $d_i = \deg f_i$, $D = \max d_i$, and ${\bf x}\in\bC^n$ such that
$J_f({\bf x})$ is invertible, then
$$\gamma(f,{\bf x}) \leq \frac{\mu(f,{\bf x}) D^{3/2}}{2 \|{\bf x}\|_1}.$$

An upper bound on $\gamma(f,{\bf x})$ also exists for systems of polynomial-exponential equations \cite{hauenstein2011certifying}.
A system is polynomial-exponential if it is polynomial in both the
variables $x_{1},\dots,x_{n}$ and finitely many exponentials of the
form $e^{a x_{i}}$ where $a\in\mathbb{C}$.
Many standard functions such as $\sin()$, $\cos()$, $\sinh()$, and $\cosh()$
can be formulated as systems of polynomial-exponential functions since they
are indeed polynomial functions of $e^{a x}$ for~suitable~$a\in\mathbb{C}$.

\subsection{Real Solutions}

The above theorem provides a bound on the distance between
an approximate solution ${\bf x}$ and its associated solution ${\bf z}$,
namely $2\beta(f,{\bf x})$.  Apart from certifying that ${\bf x}$ is
indeed an approximate solution, one often wants to prove additional
information about ${\bf z}$.  The theory of Newton-invariant sets \cite{NewtonInvariant}
provides one approach to this problem.  One particular
Newton-invariant set of particular interest is $\bR^n$ when $N_f$ defines a real map,
that is, $N_f({\bf y})\in\bR^n$ for ${\bf y}\in\bR^n$,
which was first observed in \cite{2010arXiv1011.1091H}.  In this case, one is able
to certifiably determine if ${\bf z}\in\bR^n$ or ${\bf z}\in\bC^n\setminus\bR^n$
given any approximate solution ${\bf x}$ associated with ${\bf z}$.

\newpage

The Newton iteration corresponding to a potential energy function $V$ is a real map
if $V({\bf x})$ is real for all real ${\bf x}$.  Therefore, one is able
to certifiably determine if an associated solution is real or nonreal.
This reality test and other $\alpha$-theoretic computations are
implemented in the software {\tt alphaCertified} \cite{2010arXiv1011.1091H},
which we describe next.

\subsection{Certified Region and Basins of Attraction}
The basin of attraction of each minimum is an important quantity in potential energy landscape
studies since the sum of the volumes of all the basins of attraction 
is related to the entropy of the system.
There is a crucial difference between the basins of attraction and the certifiable region of the same minimum.
The basin of attraction of a minimum cannot overlap with that of another minimum,
and aside from boundaries the union of all the basins of attraction covers the whole configuration space.
However, the certified region of a minimum is contained within the basin of attraction
of the minimum. Hence, the sum of the certifiable regions is generally a subset
of this space.

\subsection{{\tt alphaCertified}}

The software {\tt alphaCertified} \cite{2010arXiv1011.1091H} performs
computations related to $\alpha$-theory.  The input system must be presented exactly
with rational coefficients.  When the internal computations
are performed using exact rational arithmetic via GMP \cite{GMP},
all results are rigorous
and thus provide a \textit{mathematical proof of the computed results}.
This framework provides an alternative to other analytic or symbolic computations for
yielding computational proofs.  Since each solution can be independently certified,
the procedures are parallelizable.

From the exact input system, which is either a system of polynomial or polynomial-exponential functions,
the Jacobian is constructed exactly.
For these systems, the value of $\gamma$
is bounded above, as discussed in \S~\ref{Sec:SpecialNonlinear},
eliminating the need to compute the higher-order derivatives.
Hence, $\alpha$ is also bounded above.

Since the magnitude of the rational numbers can increase
during a sequence of exact computations, {\tt alphaCertified} also permits the
use of arbitrary precision floating-point arithmetic via MPFR \cite{MPFR}.
When using floating-point arithmetic, the round-off errors are not explicitly controlled but can be reduced by increasing~the~precision.

\section{Examples}\label{Sec_Examples}
In this section, we provide several other example systems illustrating
many possible scenarios in which a conventional numerical
method may face difficulty in obtaining numerical approximates.  We start with a simple example of one equation
in one variable followed by a few other examples exhibiting different numerical issues.

\subsection{An Illustrative Example}

For the system $f(x)= x^4-1$, we demonstrate the explicit computation of $\alpha, \beta$ and $\gamma$.
Since $f'(x) = 4x^3$ is zero if and only if $x = 0$ and $f(0)\neq0$, we can safely assume that $x\neq0$.
Thus, $\beta(f,x) = |x - x^{-3}|/4$.  Now, the term $D^k f(x)$ in $\gamma(f,x)$
is simply the $k$-th derivative of $f$ at $x$, namely $f^{(k)}(x)$.
Since $f$ has degree $4$, we only need to take the maximum over $k = 2,3,4$ to compute $\gamma(f,x)$.
It is easy to show that the maximum is attained at $k = 2$ with
$\gamma(f,x) = 3|x^{-1}|/2$.  Thus, $\alpha(f,x) = 3|1 - x^{-4}|/8$~for~$x\neq0$.

For $x=2.5$, we have $\alpha(f,2.5) = 0.3654$ and thus $x=2.5$ cannot be certified
as an approximate solution.  It is indeed outside of all of the
quadratic convergence basins. However, at the point $x=1.1$, $\alpha(f,1.1) = 0.11887$.
Thus, $x=1.1$ is certifiably an approximate solution of $f = 0$. In this case,
we know that the associated solution is $z = 1$, and Table \ref{tab:I} confirms
the quadratic convergence for a few iterations.

\begin{figure}[ht]
\begin{center}
\includegraphics[trim=4cm 6cm 4cm 4cm,clip,scale=0.34]{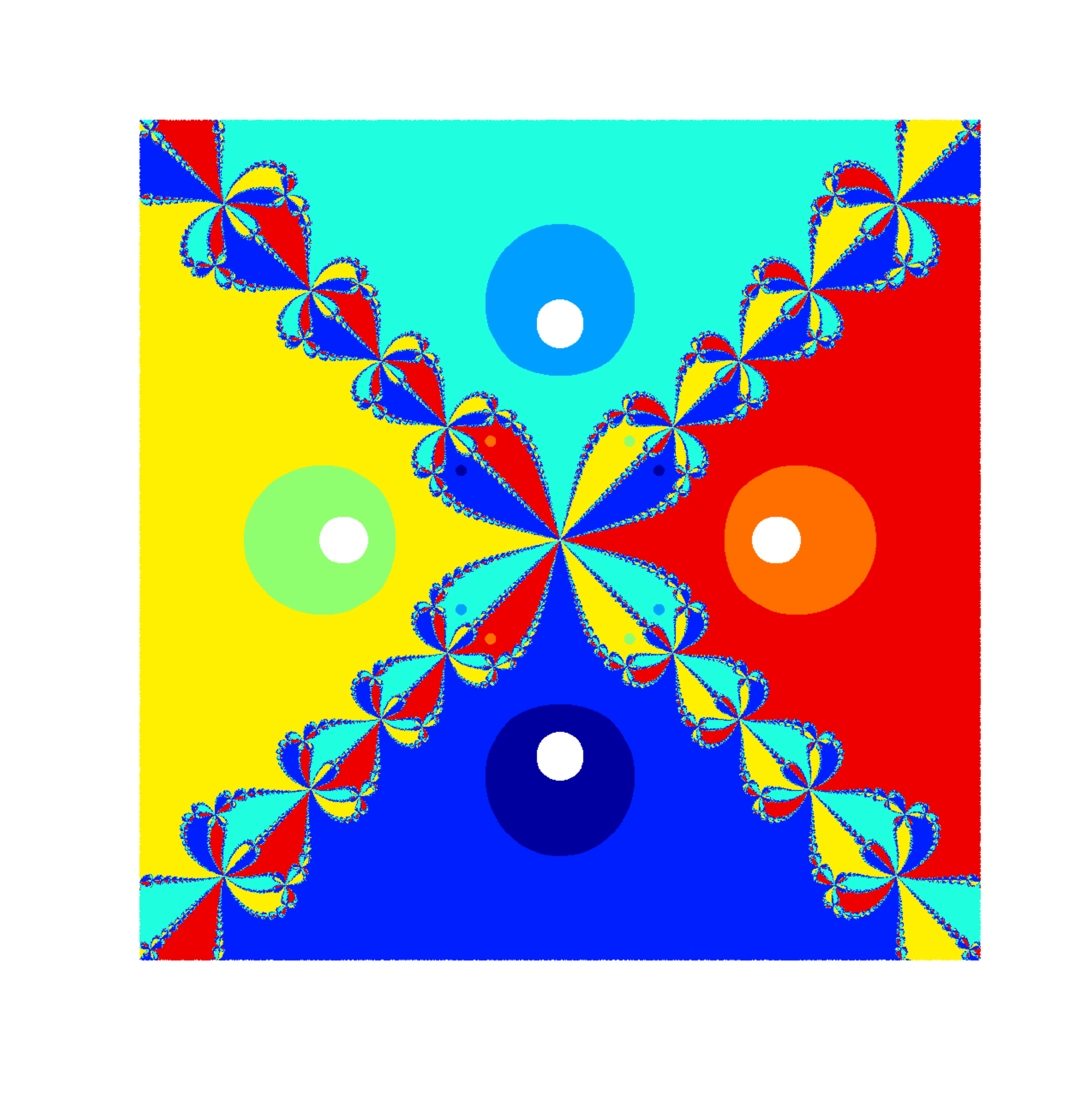}
\end{center}
\caption{Basins of convergence for Newton's method applied to $x^4 - 1$ starting at $a + b\sqrt{-1}$ for $-2 \leq a,b \leq 2$.}
\label{Fig:Z4-1}
\end{figure}

Figure~\ref{Fig:Z4-1} plots the basins of convergence starting at
points $a+bi$ for $-2\leq a,b,\leq 2$ of Newton's method applied to $f$.
In this plot, the white areas are the certifiable quadratic convergence basins
which lie inside of the respective quadratic convergence basins.
These quadratic convergence basins lie inside of the respective linear convergence basins
with chaotic behavior separating the linear convergence~basins.
The structure is similar to that observed for convergence of alternative
optimisation algorithms for atomic clusters in previous work \cite{wales92,wales93d,Asenjo13}.

\begin{table}
\caption{{\footnotesize Convergence to $z = 1$ for $f(x) = x^4-1$ starting at $x = 1.1$.}}\label{tab:I}
{\footnotesize
\begin{tabular}{c|c|c|c|c|c}
$k$ & 1 & 2 & 3 & 4 & 5\\
\hline
$-\log_{10}\left(\|N_{f}^{k}(x)-z\|\right)$ & 1.89 & 3.62 & 7.06 & 13.94 & 27.70\\
\hline
$-\log_{10}\left(\|x-z\|/2^{2^k-1}\right)$  & 1.30 & 1.90 & 3.11 & 5.52 & 10.33
\end{tabular}
}
\end{table}

\subsection{Sensitivity to Perturbations}

\begin{figure}[ht]
\begin{center}
\includegraphics[trim=3cm 6cm 0cm 6cm,clip,scale=0.51]{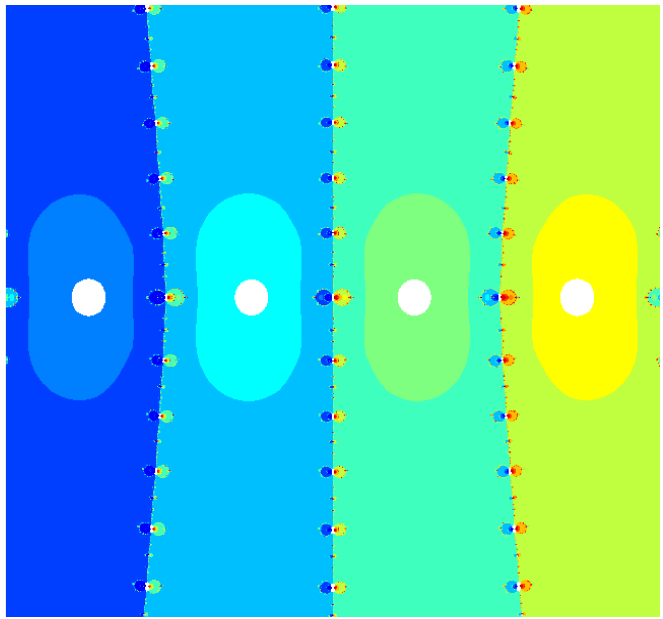}
\end{center}
\caption{Basins of convergence for Newton's method applied to 
the Wilkinson polynomial for starting at $a + b\sqrt{-1}$ for $8.5 \leq a \leq 12.5$ and
$-2 \leq b \leq 2$.}
\label{Fig:Wilkinson}
\end{figure}

Solutions of some polynomial equations are highly sensitive to perturbations.
Here, we consider the celebrated Wilkinson polynomial \cite{Wilkinson63}, the $20^{\rm th}$ degree defined by
\begin{equation}
W(x) = \prod_{j=1}^{20} (x-j) = 0.
\end{equation}
It is easy to see that the polynomial has $20$ solutions, namely $x=1,...,20$. Figure~\ref{Fig:Wilkinson}
plots the basins of convergence starting at points $a+bi$ for $8.5 \leq a \leq 12.5$
and $-2 \leq b \leq 2$ of Newton's method applied to $f$.
In this plot, the white areas are the certifiable quadratic convergence basins,
which lie inside the respective quadratic convergence basins,
separated by
chaotic behavior.

\begin{figure}[ht]
\begin{center}
\includegraphics[trim=3cm 6cm 0cm 6cm,clip,scale=0.51]{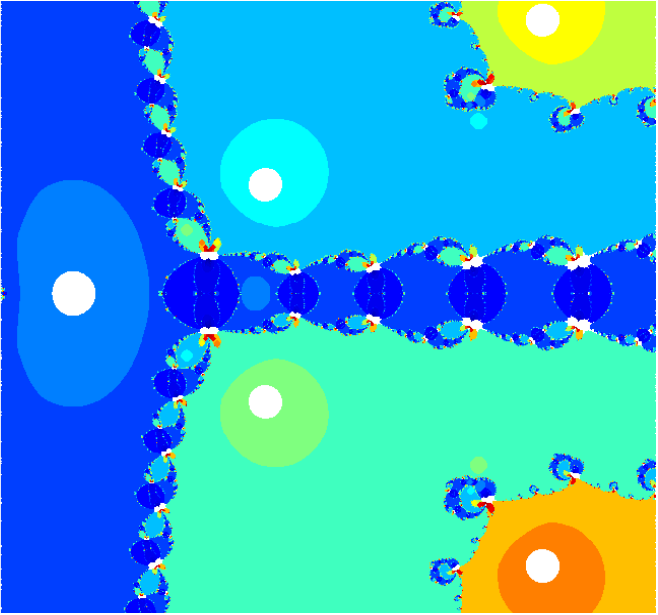}
\end{center}
\caption{Basins of convergence for Newton's method applied to perturbed Wilkinson polynomial for starting at $a + b\sqrt{-1}$ for $8.5 \leq a \leq 12.5$ and
$-2 \leq b \leq 2$.}
\label{Fig:WilkinsonPerturb}
\end{figure}

Upon expansion, the Wilkinson polynomial is
{\footnotesize
\begin{eqnarray}
W(x) &=& 2432902008176640000 - 8752948036761600000 \,x \nonumber \\
& & + 13803759753640704000 \, x^{2} - 12870931245150988800 \, x^{3} \nonumber \\
& & + 8037811822645051776 \, x^{4} - 3599979517947607200\, x^{5} \nonumber \\
& & + 1206647803780373360\, x^{6} - 311333643161390640\, x^{7} \nonumber \\
& & + 63030812099294896\, x^{8} - 10142299865511450\, x^{9} \nonumber \\
& & + 1307535010540395\, x^{10} - 135585182899530 \, x^{11} \nonumber \\
& & + 11310276995381\, x^{12} - 756111184500\, x^{13} \nonumber \\
& & + 40171771630\, x^{14} - 1672280820\, x^{15} \nonumber \\
& & + 53327946\, x^{16} - 1256850\, x^{17} \nonumber \\
& & + 20615\, x^{18} - 210 \, x^{19} + x^{20}.
\end{eqnarray}}
Wilkinson showed that even if we change the coefficient of the monomial $x^{19}$ in the above equation
from $-210$ to $-210 - 2^{-23}$, a computer with $30$-bit floating point precision would not be able to
distinguish the two numbers.
Hence, even with this change a numerical solver will give the same $20$ solutions as before.
However, the perturbed system evaluated at $x=20$ is $-2^{-23}\times20^{19} = -6.25\times 10^{17}$, i.e., $x=20$ is no longer
a solution of the equation. In fact, the solutions of the perturbed system are approximately
$$
\begin{array}{l}
1, 2, 3, 4, 5, 6.00001, 6.99970, 8.00727, 8.91725, \\
10.09527\pm \, 0.64350i, 11.79363 \pm \, 1.65233i, \\
13.99236 \pm \, 2.51883i, 16.73074 \pm \, 2.81262i, \\
19.50244 \pm \, 1.94033i, 20.84691 \end{array}.
$$
Figure~\ref{Fig:WilkinsonPerturb}
plots the basins of convergence starting at points $a+bi$ for $8.5 \leq a \leq 12.5$
and $-2 \leq b \leq 2$ of Newton's method applied to the perturbed polynomial.
In this plot, the white areas are the certifiable quadratic convergence basins
which lie inside of the respective quadratic convergence basins,
separated by chaotic behavior.

After approximating all of the solutions to the perturbed system to
roughly $28$ digits, we used {\tt alphaCertified} to prove
that this perturbed system indeed has $20$ distinct roots, only $10$ of which are real.

\subsection{Close Roots}

The $n^{\rm th}$ Chebyshev polynomial of the first kind has $n$
roots between $-1$ and $1$.  These roots, called Chebyshev nodes, are
located at $x_i = \cos\left[(2i-1)\pi/2n\right]$ for $i = 1,\dots,n$.
We can use this example to demonstrate how small perturbations in a
numerical approximation can change the root that Newton's method converges to.
This chaotic behavior can be avoided using certification.

\begin{figure}[ht]
\begin{center}
\includegraphics[trim=4cm 6cm 4cm 4cm,clip,scale=0.34]{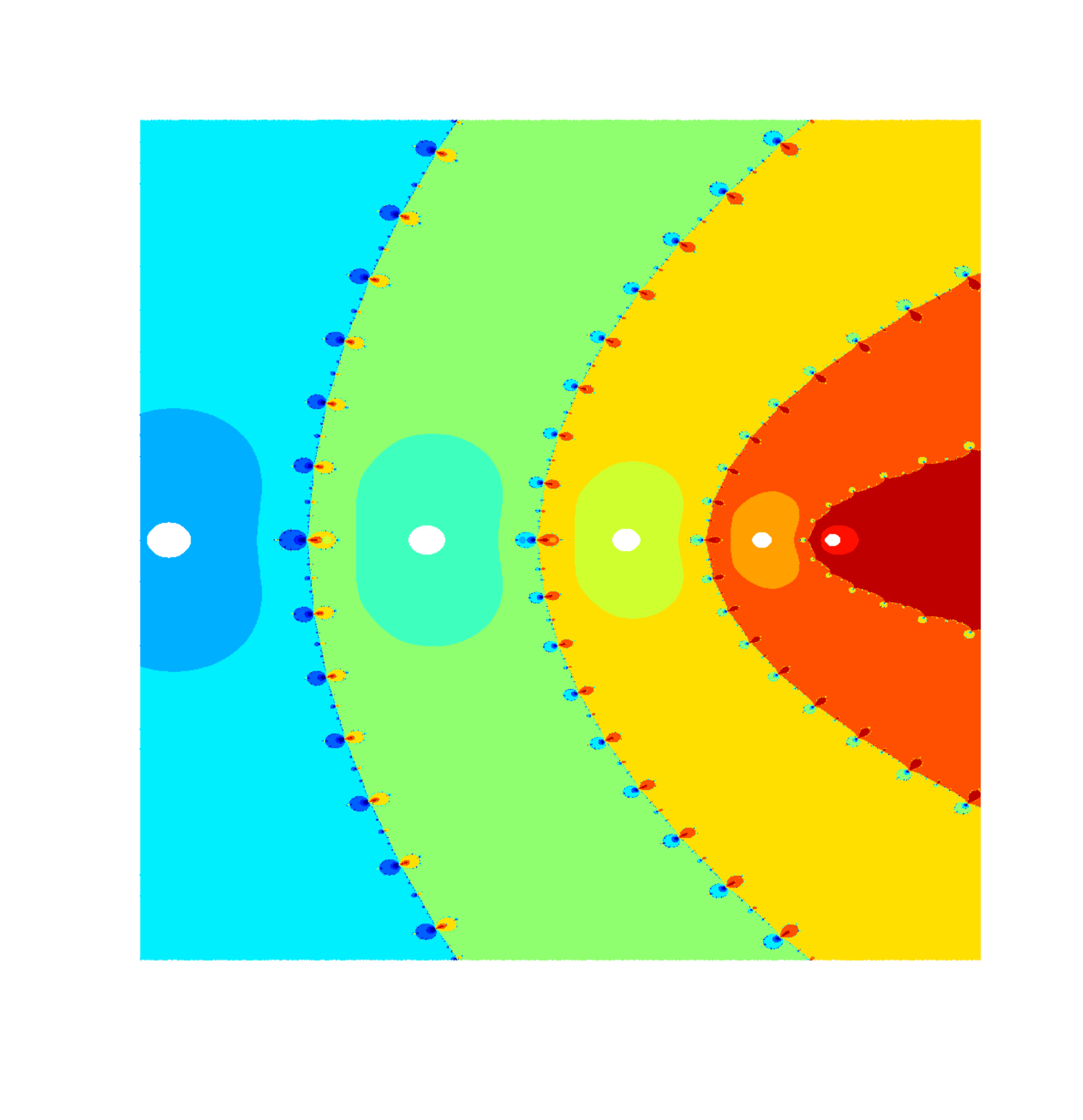}
\end{center}
\caption{Basins of convergence for Newton's method applied to $20^{th}$ Chebyshev polynomial
starting at $a + b\sqrt{-1}$: White areas are the certified region of quadratic convergence which lie inside the respective quadratic convergence basins.}
\label{Fig:Cheby}
\end{figure}

Figure \ref{Fig:Cheby} plots the basins of convergence starting at
points $a+bi$ for $0.75\leq a\leq 1.05$ and $-0.2 \leq b \leq 0.2$
of Newton's method applied to the $20^{\rm th}$ Chebyshev polynomial,
namely $f(x) = \cos(20\cos^{-1} x)$.
The white areas in this plot are the certifiable quadratic convergence basins
that lie inside of the respective quadratic convergence basins.
Along the real line, i.e., for points with $b = 0$,
the quadratic convergence basins are relatively close to chaotic regions.
Moreover, in this plotted region, there is at least one point that converges
to each of the $20$ Chebyshev nodes.

To further highlight the chaotic behavior, consider the $50^{\rm th}$
Chebyshev polynomial $f(x) = \cos(50 \cos^{-1} x)$.
Table \ref{tab:II} considers selected values near each other, which converge to
various Chebyshev nodes.

\begin{table}
\caption{{\footnotesize Convergence to various roots for the $50^{\rm th}$ Chebyshev polynomial of the first kind.}}\label{tab:II}
{\footnotesize
\begin{tabular}{l|l}
\multicolumn{1}{c|}{$x^*$} & \multicolumn{1}{c}{$\displaystyle\lim_{k\rightarrow\infty} N_f^k(x^*)$~~~~} \\
\hline
0.997 & $x_2 = \cos(3\pi/100)$ \\
0.9979 & $x_3 = \cos(5\pi/100)$ \\
0.99799 & $x_5 = \cos(9\pi/100)$ \\
0.997999 & $x_6 = \cos(11\pi/100)$ \\
0.998001 & $x_6 = \cos(11\pi/100)$ \\
0.99801 & $x_9 = \cos(17\pi/100)$ \\
0.9981 & $x_1 = \cos(\pi/100)$ \\
0.998 & $x_6 = \cos(11\pi/100)$
\end{tabular}
}
\end{table}

\section{Certifying the Minima and Transition States of the 2D XY Model}\label{sec:XY_model}
In our previous paper, we certified the SPs of the M\"uller-Brown potential as well as all the known minima and transition states \cite{DoyeW02,2005JChPh.122h4105D} of the Lennard-Jones potential for 
atomic clusters of 7 to 14 atoms
using the method described above \cite{Mehta:2013zia}.
The potential of the former model consisted of exponentials while the latter consists of rational polynomials. In this Section, we choose another important model whose potential energy landscape has
attracted interest, namely the XY model (without any disorder). The XY model consists of trigonometric terms 
illustrating a different kind of model from those previously considered.
The XY model is among the simplest lattice spin models where an energy landscape approach based on stationary points of the Hamiltonian
in a continuous configuration space
is appropriate (unlike, for example, the Ising model whose configuration
space is discrete). The potential energy landscape possesses a wide range of
interesting features, and proved to be very helpful in analyzing
the characteristic structure, dynamics, and thermodynamics. The XY model also appears in many different areas in theoretical physics, especially statistical physics \cite{kosterlitz1973ordering}, where it is employed in studies of low temperature superconductivity,
superfluid helium, hexatic liquid crystals, and Josephson junction arrays. The XY model also corresponds to the lattice Landau gauge functional for a compact
$U(1)$ lattice gauge theory \cite{Maas:2011se,Mehta:2009,Mehta:2010pe,Mehta:2014jla}.
Furthermore, it corresponds to the nearest-neighbor Kuramoto model with homogeneous
frequency, where the stationary points constitute special configurations
in phase space from a non-linear dynamical systems viewpoint
\cite{acebron2005kuramoto}.

The XY model Hamiltonian reads as:
\begin{equation}
V=\frac{1}{N^d}\sum_{j=1}^{d}\sum_{\textbf{i}}[1- \cos(\theta_{\textbf{i}+\hat{\boldsymbol\mu}_j}-\theta_{\textbf{i}})],\label{eq:F_phi}
\end{equation}
where $d$ is the dimension of a lattice, $\hat{\boldsymbol\mu}_j$ is the $d$-dimensional unit vector in the $j$-th direction,
i.e.~$\hat{\boldsymbol\mu}_1=(1,0,\ldots,0)$, $\hat{\boldsymbol\mu}_2=(0,1,0,\ldots,0)$, etc.,
$\textbf{i}$ stands for the
lattice coordinate $(i_{1},\dots,i_{d})$, and the sum over ${\textbf{i}}$ represents a sum over all $i_{1},\dots,i_{d}$ each
running from $1$ to $N$, and each $\theta_{\textbf{i}}\in(-\pi,\pi]$.
Hence $d$ is the dimension of the lattice, and $N$ is the number of sites for each dimension,
so the number of $\theta$ values required to specify the configuration is $N^d$.
The boundary conditions are given by $\theta_{\textbf{i}+N\hat{\boldsymbol\mu}_j}=(-1)^{k}\theta_{\textbf{i}}$ for
$1\le j\le d$,
where $N$ is the total number of lattice sites in each dimension,
with $k=0$ for periodic boundary conditions (PBC) and $k=1$ for
anti-periodic boundary conditions (APBC). With PBC there is a global
degree of freedom leading to a one-parameter family of solutions,
as all the equations are unchanged under $\theta_{\textbf{i}}\to\theta_{\textbf{i}}+\alpha,\forall \textbf{i}$,
where $\alpha$ is an arbitrary constant angle, due to the global O($2$) symmetry. The global symmetry can be removed
by fixing one of the variables to zero: $\theta_{(N,N,\ldots,N)}=0$.
In the present contribution, we certify all the available solutions found using the numerical
eigenvector-following method implemented in our OPTIM program~\cite{Mehta-Hughes-Schroeck-Wales2013, Mehta-Hughes-Kastner-Wales-toappear}. The solutions include minima, maxima and
saddles of all the possible indices.

We can convert the gradient of $V$, $\nabla V$, to a polynomial system by defining
$s_{\textbf{i}} = \sin(\theta_{\textbf{i}})$ and $c_{\textbf{i}} = \cos(\theta_{\textbf{i}})$
and adding the Pythagorean identities $s_{\textbf{i}}^2 + c_{\textbf{i}}^2 - 1 = 0$.
For $d = 2$ and $N = 4,\dots,10$, Tables~\ref{tab:PBC} and~\ref{tab:APBC} lists the number
of points and the average time required to perform the certification.
The difference in time is due to the reduction when fixing the
variable $\theta_{(N,N)}$ in the PBC case.  The triangle inequality based on
the maximum value of $\beta$ and the minimum pairwise distance of the approximations
yield an {\em a posteriori} verification that all of these points correspond
to distinct solutions.  Moreover, the numerical approximations for the
known stationary points just need to be approximated correct to ten digits
to have certifiable approximate solutions. The values of $2\beta$ yields how close a numerical approximation is to the actual solution. The values of $\gamma$ yields the size of the certified region around a numerical approximation. From the tables, we learn that the sizes of the certified regions remain fairly constant when increasing $N$.

\begin{table}
\caption{{\footnotesize Summary of $\alpha$, $\beta$, and $\gamma$ for PBC with $d = 2$.}}\label{tab:PBC}
{\scriptsize
\begin{tabular}{c|c|c|c|c|c|c}
  & number & average & maximum &  & maximum & minimum \\
  & of & time & upper bound & maximum & upper bound & pairwise \\
$N$ & points & (sec.) & of~$\alpha(f_N,\cdot)$ & $\beta(f_N,\cdot)$ & of~$\gamma(f_N,\cdot)$ & distance \\
\hline
$4$ & $180$ & $0.02$ & $1.39\cdot10^{-9}$ & $3.84\cdot10^{-14}$ & $3.63\cdot10^{4}$ & $0.16$ \\
$5$ & $25913$ & $0.06$ & $1.63\cdot10^{-5}$ & $7.74\cdot10^{-12}$ & $2.10\cdot10^{6}$ & $0.28$ \\
$6$ & $52140$ & $0.18$ & $6.12\cdot10^{-4}$ & $3.50\cdot10^{-11}$ & $1.75\cdot10^{7}$ & $0.22$\\
$7$ & $72207$ & $0.43$ & $1.54\cdot10^{-5}$ & $4.25\cdot10^{-12}$ & $3.63\cdot10^{6}$ & $0.66$\\
$8$ & $87889$ & $1.00$ & $2.76\cdot10^{-6}$ & $2.15\cdot10^{-12}$ & $1.28\cdot10^{6}$ & $0.90$ \\
$9$ & $106383$ & $1.89$ & $8.77\cdot10^{-7}$ & $1.19\cdot10^{-12}$ & $7.39\cdot10^{5}$ & $1.30$\\
$10$ & $121164$ & $3.51$ & $5.11\cdot10^{-6}$ & $2.43\cdot10^{-12}$ & $2.10\cdot10^{6}$ & $1.97$ \\
\end{tabular}
}
\end{table}

\begin{table}
\caption{{\footnotesize Summary of $\alpha$, $\beta$, and $\gamma$ for APBC with $d = 2$.}}\label{tab:APBC}
{\scriptsize
\begin{tabular}{c|c|c|c|c|c|c}
  & number & average & maximum &  & maximum & minimum \\
  & of & time & upper bound & maximum & upper bound & pairwise \\
$N$ & points & (sec.) & of~$\alpha(f_N,\cdot)$ & $\beta(f_N,\cdot)$ & of~$\gamma(f_N,\cdot)$ & distance \\
\hline
$4$ & $542$ & $0.02$ & $3.35\cdot10^{-9}$ & $1.61\cdot10^{-13}$ & $2.17\cdot10^{4}$ & $0.56$ \\
$5$ & $26827$ & $0.07$ & $3.24\cdot10^{-6}$ & $2.29\cdot10^{-12}$ & $1.84\cdot10^{6}$ & $0.25$\\
$6$ & $49956$ & $0.19$ & $3.02\cdot10^{-6}$ & $2.52\cdot10^{-12}$ & $1.34\cdot10^{6}$ & $0.47$\\
$7$ & $64666$ & $0.46$ & $3.06\cdot10^{-7}$ & $1.44\cdot10^{-12}$ & $5.59\cdot10^{5}$ & $0.35$\\
$8$ & $79402$ & $1.07$ & $1.03\cdot10^{-6}$ & $9.48\cdot10^{-13}$ & $1.08\cdot10^{6}$ & $0.62$\\
$9$ & $99461$ & $1.96$ & $1.14\cdot10^{-7}$ & $5.53\cdot10^{-13}$ & $4.11\cdot10^{5}$ & $0.14$\\
$10$ & $110702$ & $3.74$ & $1.98\cdot10^{-6}$ & $8.71\cdot10^{-13}$ & $2.27\cdot10^{6}$ & $2.00$ \\
\end{tabular}
}
\end{table}

To test if our set up finds singular solutions, we included some singular solutions in the list of SPs for smaller values of $N$. alphaCertified correctly found out all the singular solutions. The data is summarized in Table~\ref{tab:Sing}.

\begin{table}
\caption{{\footnotesize Number of singular solutions identified for PBC and APBC with $d = 2$.}}\label{tab:Sing}
{\footnotesize
\begin{tabular}{c|c|c}
$N$ & PBC & APBC \\
\hline
4 & 9 & 22 \\
5 & 2 & 5  \\
6 & 6 & 14 \\
\end{tabular}
}
\end{table}

Using $\alpha$-theory, one can also certifiably determine the index.
This certificate would protect against a true solution having a small eigenvalue
that changes sign when using an approximation of the solution.
Due to complications
of expanding determinants for large matrices, we use a eigenvalue vector/value
formulation at the expense of increasing the size of the system.
The resulting system is of the form
$$\left[\begin{array}{c} \nabla V \\ {\bf H}(V) {\bf v} - \lambda {\bf v} \end{array}\right] = {\bf 0}$$
where ${\bf H}(V)$ is the Hessian of $V$.  Since ${\bf v}$ is defined up to scaling, we
dehomogenize using a sufficiently random patch and then convert
to a polynomial system with the same subsitution as above.

\section{Conclusions}\label{Sec_Conclusions}

Solving nonlinear equations is one of the most frequently arising mathematical
problems in theoretical chemistry, e.g., in finding minima and transition
states of a potential energy function or finding steady states of chemical rate
equations, etc.
It is customary to resort to a numerical method to solve such systems, aside
from the rare instances when the equations can be solved exactly. For a
numerical method, a solution of a given system means a numerical approximation
of an actual solution. It is quite likely that such a numerical
approximation is a nonsolution of the system, i.e., it may lie in the linear
convergence basin or in a chaotic region, instead of the quadratic region of
convergence. In many cases, such false numerical approximations may leave us
with fundamentally different, and incorrect conclusions. In our previous
paper \cite{Mehta:2013zia}, we employed Smale's $\alpha$-theory which \textit{certifies} if a
given numerical approximation is in the quadratic convergence region of an
actual solution of the system. In the present work, we have elaborated the
mathematical concepts and a related software called alphaCertified. We then used
the certification procedure to explore the solution space of a simple problem,
$x^4-1 = 0$, where $x\in \mathbb{C}$, which already shows several regions where
a numerical method may fall into chaotic regions. We also showed the certified
regions of the known four solutions of this system. Then, we picked two
celebrated examples, the Wilkinson polynomial and a Chebyshev polynomial of the
first kind. For both systems, we explored the solution spaces and identified
the certified regions as well as regions of linear convergence and chaotic
convergences. Finally we considered the XY model in two dimensions without
disorder. There, we took the already known numerical approximations of the
saddles of all the possible indices, for the lattice sizes up to $10^2$ and
tried to certify the solutions. We chose this system to demonstrate the
applicability of the $\alpha$-theory to systems that are not in the polynomial
form. In these systems, the higher the lattice dimension is the higher the
required accuracy from a numerical method, otherwise the numerical
approximations do not fall into the quadratic convergence region, as for the
Lennard-Jones (LJ) potential \cite{Mehta-Hughes-Schroeck-Wales2013}. 
We used alphaCertified to refine the numerical
approximations whenever they were not in the quadratic convergence regions of
the corresponding solutions. We also observe that, in contrast to the
LJ case \cite{Mehta:2013zia}, for the XY model, the size of the quadratic convergence
region (or the certified region) remains fairly constant as $N$ increases. One
needs to be careful here because this difference may also arise if the
technique used to find the SPs did not find the more ill-conditioned solutions
(the ones with smaller quadratic convergence basins).

We anticipate that the certification method presented in this paper will turn out to be a standard tool to verify the numerical SPs obtained from various numerical methods. Another possible 
application is to make rigorous statements about SPs random potential surfaces, 
such as the ones studied in statistical physics and cosmology, 
as well as in pure mathematics \cite{Mehta:2013fza,Greene:2013ida}.

\section{Acknowledgement}
DM and JDH were supported by DARPA Young Faculty Award.
JDH was also supported by the National Science Foundation through DMS-1262428.
DJW and DM gratefully acknowledge support from the EPSRC~and~the~ERC. \vspace{-0.75cm}

%\bibliographystyle{jcp}
%\bibliography{bibliography_NPHC_NAG}

\end{document}